\documentclass[english,12pt]{article}
\usepackage{array}
\usepackage{graphicx}
\usepackage{amssymb}
\usepackage{comment}  
\usepackage{tikz}
\usetikzlibrary{shapes.geometric}
\usetikzlibrary{shapes.misc}
\usepackage[english]{babel}
\usepackage{amsmath}
\usepackage{multirow}
\usepackage{prettyref}
\usepackage{babel}
\usepackage{units}
\usepackage{bbm, dsfont}
\usepackage[latin1]{inputenc}
\usepackage{amsfonts}
\usepackage{amssymb}
\usepackage{babel}
\usepackage{color}
\usepackage{cite}
\def\@fmsl@sh#1#2#3{\m@th\ooalign{$\hfil#1\mkern#2/\hfil$\crcr$#1#3$}}
 \def\eq#1\en{\begin{equation}#1\end{equation}}
\def\s[#1,#2]{[#1\stackrel{\star}{,}#2]}
\def\sx[#1,#2]{[#1\stackrel{\star_{x}}{,}#2]}

\tikzset{cross/.style={cross out, draw=black, minimum size=20*(#1-\pgflinewidth), inner sep=0pt, outer sep=0pt},
cross/.default={1pt}}

\newcommand{\nc}{\newcommand}
\nc{\beq}{\begin{equation}}
\nc{\eeq}{\end{equation}}
\nc{\beqa}{\begin{eqnarray}}
\nc{\eeqa}{\end{eqnarray}}

\def\bc{\begin{center}}
\def\ec{\end{center}}

\def\to{\rightarrow}

\def\gsim{\mathrel{\mathpalette\atversim>}}

\def\bc{\begin{center}}
\def\ec{\end{center}}

\def\gsim{\mathrel{\rlap{\lower4pt\hbox{\hskip1pt$\sim$}}

    \raise1pt\hbox{$>$}}}       

\def\gsim{\mathrel{\rlap{\lower4pt\hbox{\hskip1pt$\sim$}}
    \raise1pt\hbox{$>$}}}       

\begin{document}
\makeatletter
\def\fmslash{\@ifnextchar[{\fmsl@sh}{\fmsl@sh[0mu]}}
\def\fmsl@sh[#1]#2{%
  \mathchoice
    {\@fmsl@sh\displaystyle{#1}{#2}}%
    {\@fmsl@sh\textstyle{#1}{#2}}%
    {\@fmsl@sh\scriptstyle{#1}{#2}}%
    {\@fmsl@sh\scriptscriptstyle{#1}{#2}}}
\def\@fmsl@sh#1#2#3{\m@th\ooalign{$\hfil#1\mkern#2/\hfil$\crcr$#1#3$}}
\makeatother

\thispagestyle{empty}
\begin{titlepage}
\boldmath
\begin{center}
  \Large {\bf  Gravitational Radiation in Quantum Gravity}
    \end{center}
\unboldmath
\vspace{0.2cm}
\begin{center}
{  {\large Xavier Calmet}\footnote{x.calmet@sussex.ac.uk}$^{,a,b}$},
{  {\large Basem Kamal El-Menoufi}\footnote{b.elmenoufi@sussex.ac.uk}$^{,a}$},
{  {\large Boris Latosh}\footnote{b.latosh@sussex.ac.uk}$^{,a,c}$} {\large and} 
{  {\large Sonali Mohapatra}\footnote{s.mohapatra@sussex.ac.uk}$^{,a}$}
 \end{center}
\begin{center}
$^a${\sl Department of Physics and Astronomy, 
University of Sussex, Brighton, BN1 9QH, United Kingdom
}\\
$^b${\sl PRISMA Cluster of Excellence and Mainz Institute for Theoretical Physics, Johannes Gutenberg University, 55099 Mainz, Germany }\\
$^c${\sl Dubna State University,
Universitetskaya str. 19, Dubna 141982, Russia}
\end{center}
\vspace{5cm}
\begin{abstract}
\noindent
The effective field theory of quantum gravity generically predicts non-locality to be present in the effective action, which results from the low-energy propagation of gravitons and massless matter. Working to second order in gravitational curvature, we reconsider the effects of quantum gravity on the gravitational radiation emitted from a binary system. In particular, we calculate for the first time the leading order quantum gravitational correction to the classical quadrupole radiation formula which appears at second order in Newton's constant.
\end{abstract}  
\vspace{5cm}
\end{titlepage}



\newpage
\section{Introduction}

The aim of this work is to extend the study of quantum gravitational corrections to gravitational radiation initiated in \cite{Calmet:2017rxl,Calmet:2018qwg} using effective theory techniques to treat quantum gravity in a model independent way. In previous papers \cite{Calmet:2017rxl,Calmet:2018qwg} the authors focused on the production of new massive modes present in the effective action \cite{Calmet:2014gya}. We expand on the previous analyses and calculate for the first time the genuine quantum gravitational correction to the quadrupole radiation formula first developed by Einstein. While the effect is way too small to be observable by the current gravitational wave observatories and thus has no impact for the recent gravitational wave observations \cite{Abbott:2016blz,TheLIGOScientific:2017qsa}, our work offers a proof of principle that genuine calculations within quantum gravity at energies below the Planck mass  are possible, even though we do not yet have a fully satisfactory ultra-violet complete theory of quantum gravity.

We follow the approach introduced by Weinberg \cite{Weinberg}  in the 70's and further developed by others \cite{Bar1984,Bar1985,Donoghue:1994dn}. The main benefit of the effective theory approach is its ability to separate out low-energy dynamics from the unknown ultra-violet physics associated with the completion of quantum gravity. Quantum general relativity has indeed a poor ultra-violet behavior, i.e. it is non-renormalizable, yet the unknown physics is solely encoded in the Wilson coefficients of the most general diffeomorphism invariant {\em local} Lagrangian. When the Wilson coefficients are measured, any observable computed in the effective theory is completely determined to any desired accuracy in the effective field theory expansion. More interesting are the contributions induced by long-distance propagation of massless (light) degrees of freedom. The latter comprise reliable and parameter-free, and thus model independent, predictions of quantum gravity since, by the very nature of the effective field theory, any ultra-violet completion must reproduce these results at low energies.

In this paper we revisit the long-distance limit of quantum gravity and the signatures thereof on the gravitational radiation emitted from binary systems. As we shall describe below, quantum corrections are encoded in a covariant effective action organized as an expansion in gravitational curvatures. Moreover, low-energy quantum effects manifest in the effective action via a covariant set of non-local operators. The three phases of the binary evolution will be affected by quantum corrections. Thanks to advances in infrared quantum gravity \cite{Bar1984,Bar1985,Bar1987,Bar1990,Donoghue:2014yha,Cod2015}, we could in principle determine the modified fate of each phase since the effective action retains the non-linear structure of the field equations. Nevertheless, to obtain analytic insight we only focus on the leading quantum corrections to the quadrupole radiation of general relativity. It is important to keep in mind that the initial stage of a coalescence process is the only part one can study with analytical tools.

We shall define two schemes to treat quantum corrections. The first is {\em non-perturbative}, in the sense that higher-derivative terms in the equations of motion are considered on the same footing as those of general relativity. We focus on the massive spin-2 sector and show that the propagator has a multi-sheet complex structure \cite{Calmet:2017omb}, which arises due to the logarithmic non-analyticity in the equations of motion. The imaginary part of the complex poles causes the massive spin-2 field to exhibit a Yukawa suppression in the far-field region. The second treatment is {\em perturbative} and aligns naturally with the power-counting of the effective theory. Namely, we look for small corrections to the lowest-order general relativity result, i.e. quadrupole radiation, and solve the equations of motion by iteration. This is the genuine quantum gravitational correction discussed early and the main new result of this paper. In the latter scheme, the correction to the spin-2 sector is a traveling wave at the speed of light, but the amplitude falls off faster than $1/r$.

Before we proceed, it is crucial to describe the physical content of our results. All our analysis is performed on the linear weak-field level, but general relativity and the associated quantum corrections are inherently non-linear. This distinction is crucial when one deviates from pure general relativity. Indeed, it was shown in \cite{Calmet:2017qqa} that an eternal Schwarzschild black hole is a solution to the {\em full} non-linear quantum corrected theory. On the contrary and due to the breakdown of Birkhoff's theorem, the gravitational field around a non-vacuum source such as a star receives a genuine quantum correction \cite{Calmet:2017qqa}. Hence, all our results will only pertain to the inspiraling phase of mergers where the gravitational radiation is sourced by horizonless objects such as neutron stars or black holes if we think of them as objects which are not vacuum solutions but rather astrophysical objects which are still experiencing gravitational collapse \cite{Calmet:2018vuz}.

The paper is organized as follows. In Section \ref{review} we start with a brief review of the effective theory and write down the non-local corrections we shall investigate. Section \ref{local} is devoted to a quick survey of the radiation problem in local quadratic gravity. Section \ref{nonpert} and \ref{pert} treat the non-local corrections in the two different schemes described above. We conclude in Section \ref{Concs}. A careful derivation of the non-local kernel used in Section \ref{pert} is laid out in an appendix.

\section{The non-local quantum corrections}\label{review}

The effective field theory treatment of quantum gravity is by now very well understood. The initial incarnation of the effective field theory was designed mainly to compute scattering amplitudes in flat space. For example, graviton-graviton scattering can be obtained to any desired accuracy in the counting parameter of the effective theory, i.e. $(GE^2)^n$ where $E$ is the center-of-mass energy of the process. At lowest-order $\mathcal{O}(GE^2)$, one extracts vertices from the Einstein-Hilbert action and computes tree-level diagrams. At order $\mathcal{O}(GE^2)^2$, one-loop diagrams appear and the ultra-violet divergences renormalize the Wilson coefficients of the quadratic curvature action. The framework is readily extended to include matter fields. In summary, the action of the effective theory, accurate to order $(GE^2)^2$, reads\footnote{Notice that in writing this action we have employed the Gauss-Bonnet identity to get rid of the Riemann squared invariant. We also dropped a total derivative, $\Box R$, that does not provide a non-trivial Feynman rule. Also note that the power counting in $\mathcal{L}_m$ depends on the mass of the matter field.}
\begin{align}\label{EFTaction}
S_{\text{EFT}} = \int_{\mathcal{M}} \left( \frac{R}{16\pi G} + c_1 R^2 + c_2 R_{\mu\nu} R^{\mu\nu} + \mathcal{L}_m \right) \ \ .
\end{align}

To complete the effective field theory program, a measurement of the Wilson coefficients is required as per usual with any ultra-violet-sensitive quantity in quantum field theory. Unfortunately, such experimental input is not available in our case and one might question if the effective field theory is able to make any predictions. It was the point of view developed in \cite{Donoghue:1994dn} where it is shown that there exist a class of quantum corrections that comprise reliable signatures of quantum gravity. The latter appear as {\em finite} non-analytic functions in loop processes and arise directly from the low-energy propagation of virtual massless quanta. As such, these corrections are purely of infra-red origin modifying the long-distance dynamics of gravitation. A prime example is the correction to the non-relativistic Newtonian potential energy \cite{BjerrumBohr:2002kt}
\begin{align}\label{newtonian}
V_{\text{N}}(r) = -\frac{Gm_1 m_2}{r} \left( 1 + \frac{3 G (m_1 + m_2)}{r} + \frac{41}{10\pi^2} \frac{l_{P}^2}{r^2} \right) \ \ .
\end{align}

Moving ahead of scattering amplitudes, one inquires about the structure of long-distance quantum effects in the effective action. A substantial body of work has been devoted to construct the effective action of quantum gravity that encapsulates such quantum corrections. We refer the interested reader to the following articles and references therein \cite{Bar1984,Bar1985,Bar1987,Bar1990,Donoghue:2014yha,Cod2015}. Here, we merely quote the leading operators in the non-local curvature expansion 
\begin{align}\label{nonlocalaction}
	\Gamma_{\text{NL}}^{\scriptstyle{(2)}}  = - \int_{\mathcal{M}} \left[ \alpha R \ln\left(\frac{\Box}{\mu^2}\right)R + \beta R_{\mu\nu} \ln\left(\frac{\Box}{\mu^2}\right) R^{\mu\nu} + \gamma R_{\mu\nu\alpha\beta} \ln\left(\frac{\Box}{\mu^2}\right)R^{\mu\nu\alpha\beta} \right],
	\end{align}
	where $\Box := g^{\mu\nu} \nabla_\mu \nabla_\nu$. The precise values of the coefficients depend on the spin of the massless particle that runs in the loop and are listed in table (\ref{coeff1}). Non-local effective actions open the door to (re)-examine plenty of questions in gravitational physics. In this paper, we shall focus on the effect of Eq.~(\ref{nonlocalaction}) on the production of gravitational radiation from binary systems.
	
	\begin{table}
\center
\begin{tabular}{| c | c | c | c |}
\hline
 & $ \alpha $ & $\beta$ & $\gamma$  \\
 \hline
 \text{Scalar} & $ 5(6\xi-1)^2$ & $-2 $ & $2$     \\
 \hline
 \text{Fermion} & $-5$ & $8$ & $7 $ \\
 \hline
 \text{Vector} & $-50$ & $176$ & $-26$ \\
 \hline
 \text{Graviton} & $250$ & $-244$ & $424$\\
 \hline
\end{tabular}
\caption{Coefficients for different fields. Note that these coefficients have been derived by many different authors, see e.g. \cite{Birrell:1982ix,Mirzabekian:1993nz,Donoghue:1994dn,Elizalde:1995tx,Han:2004wt,Mirzabekian:1998ha,Bar1984,Bar1985,Donoghue:2014yha}. All numbers should be divided by $11520\pi^2$. Here, $\xi$ denotes the value of the non-minimal coupling for a scalar theory. All these coefficients including those for the graviton are gauge invariant. It is well known that one needs to be careful with the graviton self-interaction diagrams and that the coefficients $\alpha$ and $\beta$ can be gauge dependent, see \cite{Kallosh:1978wt}, if the effective action is defined in a naive way. For example, the numbers $\alpha= 430/(11520\pi^2)$ and $\beta=-1444/(11520\pi^2)$ for the graviton quoted in \cite{Donoghue:2014yha} are obtained using the Feynman gauge. However, there is a well-established procedure to derive a unique effective action which leads to gauge independent results \cite{Bar1984,Bar1985}. Here we are quoting the values of $\alpha$ and $\beta$ for the graviton obtained using this formalism as it guaranties the gauge independence of observables.}
\label{coeff1}
\end{table}

\section{Production of gravitational waves: local theory}\label{local}

As explained in \cite{Calmet:2017rxl,Calmet:2018qwg}, quantum gravity contains two massive wave solutions on top of the usual massless mode of general relativity. We review the results presented in \cite{Calmet:2017rxl,Calmet:2018qwg} in preparation for calculation of the leading order quantum gravitational correction to the classical quadrupole formula.  To streamline the discussion, we shall focus in this section on the local quadratic theory, i.e. Eq.~(\ref{EFTaction}). Analyzing the latter, albeit simple in nature, aids in drawing interesting parallels and contrasts when we discuss non-locality in the next section. We only consider a simple system where the two masses move in a perfectly circular orbit. 

The equations of motion are easily obtained by linearizing the field equations of Eq.~(\ref{EFTaction})
\begin{align}\label{LEOM}
&\Box\bar{h}_{\mu \nu} - \kappa^2 \Box  \Big [ \left(c_1 + \frac{c_2}{2} + c_3 \right) \partial_\mu \partial_\nu \bar{h} - \left(c_1 + \frac{c_2}{2} + c_3 \right) \eta_{\mu\nu} \Box \bar{h} + \left(\frac{c_2}{2} + 2 c_3 \right) \Box \bar{h}_{\mu \nu} \Big ] = -16 \pi G T_{\mu\nu}
\end{align}
where $\bar{h}_{\mu\nu} \equiv h_{\mu\nu} - \frac12 \eta_{\mu\nu} h$ is the trace-reduced tensor, $\kappa^2 = 32 \pi G$ and we employed the harmonic gauge. It is more convenient to perform our calculation using the trace-reduced tensor, and only at the end obtain $h_{\mu\nu}$ by subtracting off the trace. Since the pioneering work of Stelle \cite{Stelle:1977ry}, it became quite common to dispense with the higher-derivative structure of the theory by introducing massive modes in the equations of motion. These extra modes decouple from the massless spin-2 mode. Working in momentum-space, we get
\begin{align}
\bar{\mathcal{O}}_{\mu\nu}^{~~\alpha\beta}\, \bar{h}_{\alpha\beta}(k) = - 16 \pi G T_{\mu\nu}(k)
\end{align}
where
\begin{eqnarray}\label{localop}
\bar{\mathcal{O}}_{\mu\nu}^{~~\alpha\beta} &=& -\frac{k^2}{2} (\delta^\alpha_\mu \delta^\beta_\nu + \delta^\alpha_\nu \delta^\beta_\mu) 
\\ \nonumber && - \kappa^2 \left[\left(c_1 + \frac{c_2}{2} + c_3 \right) ( k^2 k_\mu k_\nu \eta^{\alpha\beta} - k^4 \eta_{\mu\nu} \eta^{\alpha\beta} ) + \left(\frac{c_2}{2} + 2 c_3 \right) \frac{k^4}{2} (\delta^\alpha_\mu \delta^\beta_\nu + \delta^\alpha_\nu \delta^\beta_\mu)\right] \ \ .
\end{eqnarray}
Revealing the massive modes requires that we project out the spin-2 and spin-0 parts of the symmetric operator
\begin{align}
\mathcal{P}_{\mu\nu}^{(2)\alpha\beta} = \frac12 \left(\theta_\mu^\alpha \theta_\nu^\beta + \theta_\nu^\alpha \theta_\mu^\beta \right) - \frac13 \theta_{\mu\nu} \theta^{\alpha\beta}, \quad\mathcal{P}_{\mu\nu}^{(0)\alpha\beta} = \frac13 \theta_{\mu\nu} \theta^{\alpha\beta},
\end{align}
where $\theta_{\mu\nu} = \eta_{\mu\nu} - k_\mu k_\nu/k^2$. In harmonic gauge, we have $k^\mu \bar{h}_{\mu\nu} = 0$ and so Eq.~(\ref{localop}) is easily rewritten as
\begin{align}\label{spinlocalop}
\bar{\mathcal{O}}_{\mu\nu}^{~~\alpha\beta} = -k^2 \left(1 + \kappa^2 \left(\frac{c_2}{2} + 2 c_3 \right) k^2 \right) \mathcal{P}_{\mu\nu}^{(2)\alpha\beta} - k^2 \left(1 + \kappa^2 \left( -3 c_1 - c_2 - c_3 \right)  k^2 \right) \mathcal{P}_{\mu\nu}^{(0)\alpha\beta} \ \ .
\end{align}
Inverting the operator yields the propagator in momentum-space
\begin{align}\label{localprop}
\bar{\mathcal{D}}_{\mu\nu}^{~~\alpha\beta} = -\frac{\left(\mathcal{P}_{\mu\nu}^{(2)\alpha\beta} + \mathcal{P}_{\mu\nu}^{(0)\alpha\beta}\right)}{k^2} + \frac{\mathcal{P}_{\mu\nu}^{(2)\alpha\beta} }{k^2 - m_2^2} + \frac{\mathcal{P}_{\mu\nu}^{(0)\alpha\beta}}{k^2 - m_0^2}
\end{align}
where we have used partial fractions to identify the masses of the spin-2 and spin-0 sectors
\begin{align}
m_2^2 =\frac{ M_P^2}{2(- c_2 - 4 c_3)}, \quad m_0^2 =\frac{M_P^2}{4(3 c_1 + c_2 + c_3)} \ \ .
\end{align}

We stress again that Eq.~(\ref{localprop}) is the propagator for $\bar{h}_{\mu\nu}$. For completeness, we can easily obtain the appropriate propagator for $h_{\mu\nu}$ by subtracting the trace of Eq.~(\ref{localprop})
\begin{align}\label{localproph}
\nonumber
\mathcal{D}_{\mu\nu}^{~~\alpha\beta} &= \bar{\mathcal{D}}_{\mu\nu}^{~~\alpha\beta} - \frac12 \eta_{\mu\nu} \eta^{\gamma\lambda} \bar{\mathcal{D}}_{\gamma\lambda}^{~~\alpha\beta}\\
 &= - \frac{\delta_\mu^\alpha \delta_\nu^\beta+ \delta_\nu^\alpha \delta_\mu^\beta - \eta_{\mu\nu} \eta^{\alpha\beta}}{2 k^2} + \frac{\mathcal{P}_{\mu\nu}^{(2)\alpha\beta} }{k^2 - m_2^2} - \frac{\mathcal{P}_{\mu\nu}^{(0)\alpha\beta}}{2(k^2 - m_0^2)}
\end{align}
which is the known result derived by Stelle \cite{Stelle:1977ry}. As emphasized, the extension of general relativity including the terms quadratic in curvature contains three mass eigentstates: a massless mode with spin-2 and two massive modes with respectively spin 2 and 0.
The massive spin-2 mode is formally a ghost while the massive spin-0 mode is healthy. However, as already explained in details in \cite{Calmet:2018qwg,Calmet:2018uub}, the massive spin-2, although it is formally a ghost, does not lead to any pathology. The effective action contains only classical fields, as the fluctuations of the graviton have been integrated out. The massive field with spin-2 can simply be seen as a field that couples with minus the Planck scale to the stress-energy tensor. It is a nothing but a repulsive force.  Notice also here that either (or both) of $m_0$ and $m_2$ could be tachyonic depending on the exact values of the Wilson coefficients. In this section, we proceed under the assumption that the masses are real. 

Given the manifest decoupling of the modes, the solution to Eq.~(\ref{LEOM}) is the direct sum of the three sectors. One can switch back to position-space and write down the solution for the trace-reduced metric perturbations, making sure to define the propagators with retarded boundary conditions
\begin{eqnarray}\label{localGreen}
\bar{h}_{\mu \nu} &= &16 \pi G \int d^4x^\prime \, G^{\text{ret.}} (x-x^\prime; 0) T_{\mu\nu}(x^\prime) \\ \nonumber &&  - 16 \pi G \int d^4x^\prime \, G^{\text{ret.}} (x-x^\prime; m_2) \mathcal{P}_{\mu\nu}^{(2)\alpha\beta} T_{\alpha\beta}(x^\prime)  \\ \nonumber && 
- 16 \pi G \int d^4x^\prime \, G^{\text{ret.}} (x-x^\prime; m_0) \mathcal{P}_{\mu\nu}^{(0)\alpha\beta} T_{\alpha\beta}(x^\prime).
\end{eqnarray}
Note that the general relativity solution is given by
\begin{align}
\bar{h}^{\text{GR}}_{\mu \nu} := 16 \pi G \int d^4x^\prime \, G^{\text{ret.}} (x-x^\prime; 0) T_{\mu\nu}(x^\prime)
\end{align}
It is important to realize that the two new terms are of the same order in $G$ as the usual solution from general relativity. These are not corrections to general relativity solutions. There are simply additional classical modes present in the action. We stress that each of these terms is a solution to their partial differential equations which are fully decoupled. We write them as a direct sum for convenience, but the reader should not get confused.

We consider our source to be a simple binary system and set the origin of the coordinates to coincide with the center-of-mass of the system
\begin{align}
T_{\mu\nu} = \sum_{i=1}^2 M_i \dot{x}_\mu \dot{x}_\nu\, \delta^{(3)}\left(\vec{x} - \vec{X}_i(\tau)\right) 
\end{align}
where a dot denotes a derivative with respect to proper time, $\tau$, and $\vec{X}_i$ is the trajectory of the mass. In the slow-velocity limit, proper time coincides with coordinate time to lowest order in velocity. We notice first that the spin-0 mode couples to the trace of the energy-momentum tensor, which is time-independent for a binary system in circular orbit. Focusing on the massive spin-2 sector, we are interested in the leading behavior in the far-zone ($|\vec{x} - \vec{x}^\prime| \approx |\vec{x}| := r $). It suffices to solve for the spatial components, i.e. $\bar{h}_{ij}$, the other metric perturbations are determined using the harmonic gauge condition. With this set-up, Eq.~(\ref{localGreen}) becomes\footnote{In writing Eq.~(\ref{resultsemifinal}) we ignored all terms proportional to the trace of the energy-momentum tensor, which is time independent for a binary in circular orbit.}
\begin{align}\label{resultsemifinal}
\bar{h}_{ij} = \bar{h}^{\text{GR}}_{ij} - 16 \pi G \int d\omega\, e^{-i\omega t} I_{ij}(\omega) \int \frac{k^2 dk d\Omega_k}{(2\pi)^3} \frac{e^{i \vec{k} \cdot \vec{x}}}{(\omega+i\epsilon)^2 - k^2 - m_2^2}
\end{align}
where
\begin{equation}\label{quadtensor}
I_{ij} (\omega)  = - \frac12 \mu  ( d \omega_s)^2 \left(\begin{array}{ccc} \delta(\omega+2\omega_s) + \delta(\omega - 2\omega_s)  & -i(\delta(\omega+2\omega_s) - \delta(\omega - 2\omega_s)) & 0\\ -i(\delta(\omega+2\omega_s) - \delta(\omega - 2\omega_s)) & -\delta(\omega+2\omega_s) - \delta(\omega - 2\omega_s) & 0  \\ 0 & 0 & 0 \end{array}\right) \ \ .
\end{equation}
In the above, $\mu$ is the reduced mass of the binary, $d$ is the orbital separation and $\omega_s$ is the orbital frequency. In Eq.~(\ref{resultsemifinal}), notice most importantly the $i\epsilon$ prescription is due to the retarded boundary conditions. The angular integrals in Eq.~(\ref{resultsemifinal}) are readily done, and the final integral over the spatial momentum depends crucially on the size of the mass compared to the orbital frequency. In the complex $k$-plane, the poles are situated at
\begin{align}\label{localpoles}
k_{\pm} = \pm \sqrt{\omega^2 - m_2^2} \pm \text{sgn}(\omega)\,i \epsilon \ \ .
\end{align}
One notices two features of the above expression. First, the poles are real (imaginary) if the mass is smaller (greater) than the frequency. Second, if the poles are real then the sign of the frequency is important in moving the poles off the real axis, which is paramount in obtaining a proper propagating wave. After a careful computation we find
\begin{align}\label{finalresultlocal}
\bar{h}_{ij} (t,r)= \bar{h}^{\text{GR}}_{ij} - 4 G \frac{\mu  ( d \omega_s)^2}{r} \left[ \theta(m_2-2\omega_s) e^{-\sqrt{m_2^2 - 4 \omega_s^2}r} Q_{ij}(t,0;0) + \theta(2\omega_s - m_2) Q_{ij}(t,r;m_2^2)\right] 
\end{align}
where we defined
\begin{equation}\label{Qtensor}
Q_{ij} (t,r;m^2)  = \left(\begin{array}{ccc} \cos \left(2\omega_s \left(t - \sqrt{1-(m/2\omega_s)^2} r\right) \right)  & \sin \left(2\omega_s \left(t - \sqrt{1-(m/2\omega_s)^2} r\right) \right) & 0\\ \sin \left(2\omega_s \left(t - \sqrt{1-(m/2\omega_s)^2} r \right) \right) & -\cos \left(2\omega_s \left(t - \sqrt{1-(m/2\omega_s)^2} r \right) \right) & 0  \\ 0 & 0 & 0 \end{array}\right) \ \ .
\end{equation}
The remaining integrals can now easily be performed. We find 
\begin{align}
h_{ij} (t,r)= h^{\text{GR}}_{ij} - 4 G \frac{\mu  ( d \omega_s)^2}{r} \left[ \theta(m_2-2\omega_s) e^{-\sqrt{m_2^2 - 4 \omega_s^2}r} Q_{ij}(t,0;0) + \theta(2\omega_s - m_2) Q_{ij}(t,r;m_2^2)\right],
\end{align}
in the far zone, where
\begin{align}
h^{\text{GR}}_{ij} : = 4 G \frac{\mu  ( d \omega_s)^2}{r} Q_{ij}(t,r;0) \ \ .
\end{align}

Comments about the above result are in place:
\begin{itemize}
\item The second term has the opposite sign in comparison to that of general relativity, which signifies the repulsive nature of the massive spin-2 sector.  This mode is {\em classically} healthy because it carries positive-definite energy. To compute the radiated power, one simply has to construct the energy-momentum tensor from the Lagrangian of the theory. Since the different modes are decoupled \cite{Stelle:1977ry}, the total energy-momentum tensor is likewise decoupled. The latter is quadratic in the field variables and so obviously the negative sign in the massive spin-2 solution does not affect the positivity of the energy.
\item Eq.~(\ref{finalresultlocal}) contains two parts. If the mass is large compared to the characteristic frequency of the system, the result is a standing wave due to the Yukawa suppression. Hence, formally no energy is transmitted to infinity. The traveling wave portion has outgoing spherical wave-fronts and is viable only if the frequency is large enough to excite the massive mode.
\item The $i\epsilon$ prescription is crucial to obtain a solution that represents a traveling wave: the position of the poles changes when the frequency flips from $\omega = 2 \omega_s$ to $\omega = - 2 \omega_s$. This takes place consistently such that all exponential factors arrange correctly and yield sinusoidal functions propagating at the correct speed appropriate for a massive wave.
\item The wave is sub-luminal and has a group velocity  
$v_g(\omega) = \sqrt{1-(m_2/\omega)^2}$, which is readily identified from the dispersion relation $k(\omega) = \omega \sqrt{1-(m/\omega)^2}$. This is precisely the relativistic velocity of a free massive particle.
\item For completeness, we can easily compute the total emitted power. We use the fact that the total energy-momentum tensor is the direct sum of the three modes and notice that the energy-momentum tensor of a massive spin-2 theory is identical to that of general relativity\footnote{Notice that this is true in general, i.e. not necessarily requiring the Pauli-Fierz tuning.}. To lowest order in the mass, we have the rate of energy loss
\begin{align}
\frac{dE_{\text{GW}}}{dt}  = \frac{32 G \mu^2 d^4 \omega_s^6}{5} \left( 1 + \theta(2\omega_s - m_2) \right) + \mathcal{O}\left(\frac{m_2}{\omega_s}\right) \ \ .
\end{align}  
where, as explained in  \cite{Calmet:2018qwg} where this equation was first derived, the first term is the power lost in the massless gravitational mode while the second term represents the power lost in the massive spin-2 mode.
\end{itemize}


\section{Quantum non-locality: Non-perturbative treatment}\label{nonpert}
We now include the non-local higher curvature corrections in the equations of motion.  Adding the non-local corrections, we find (in harmonic gauge)
\begin{align}\label{NLEOM}
\nonumber
&\Box\bar{h}_{\mu \nu} - \kappa^2 \Box \Big [ \left(c_1(\mu) + \frac{c_2(\mu)}{2} + c_3(\mu) \right) (\partial_\mu \partial_\nu - \eta_{\mu\nu} \Box) \bar{h} - \left(\alpha + \frac{\beta}{2} + \gamma \right) \frak{L}(\bar{h}),_{\mu \nu} \\
&+ \left(\alpha + \frac{\beta}{2} + \gamma \right) \eta_{\mu\nu} \Box \frak{L}(\bar{h}) + \left(\frac{c_2(\mu)}{2} + 2 c_3(\mu) \right) \Box \bar{h}_{\mu \nu} - \left(\frac{\beta}{2} + 2\gamma \right) \Box \frak{L}(\bar{h}_{\mu \nu}) \Big ] = -16 \pi G T_{\mu\nu}
\end{align}
where
\begin{align}
\frak{L}(f) := \int d^4x^\prime \frak{L}(x-x^\prime) \, f(x^\prime), \quad  \frak{L}(x-x^\prime) = \int \frac{d^4 k}{(2\pi)^4} e^{-i k (x-x^\prime)} \ln \left(\frac{-k^2}{\mu^2}\right) \ \ .
\end{align}
The non-local function, $\frak{L}(x-x^\prime)$, must be supplemented by a boundary condition to be well-defined. We impose retarded boundary conditions by sending $k^0 \to k^0 + i \epsilon$ inside the logarithm; see the discussion in the appendix. The exact form of $\frak{L}(x-x^\prime)$ is derived in Appendix~(\ref{app}), nevertheless, we will not need such an expression in this section. In fact, we wish to treat the higher-derivative terms along the same lines of the last section. We refer to this treatment as {\em non-perturbative}, and so we transform Eq.~(\ref{NLEOM}) to momentum-space and obtain the non-analytic operator
\begin{align}\label{spinnonlocalop}
\nonumber
\bar{\mathcal{O}}_{\mu\nu}^{~~\alpha\beta} &= -k^2 \left(1+ \kappa^2 \left(\frac{c_2(\mu)}{2} + 2 c_3(\mu) \right) k^2 - \kappa^2 \left(\frac{\beta}{2} + 2 \gamma \right) k^2 \ln \left(\frac{-k^2}{\mu^2}\right) \right) \mathcal{P}_{\mu\nu}^{(2)\alpha\beta} \\
&- k^2 \left(1+ \kappa^2 \left( -3 c_1(\mu) - c_2(\mu) - c_3(\mu) \right)  k^2  - \kappa^2 \left( -3 \alpha - \beta - \gamma \right)  k^2 \ln \left(\frac{-k^2}{\mu^2}\right) \right) \mathcal{P}_{\mu\nu}^{(0)\alpha\beta} 
\end{align}
whose propagator is readily constructed
\begin{align}\label{nonlocalprop}
\nonumber
\bar{\mathcal{D}}_{\mu\nu}^{~~\alpha\beta} &= \frac{\mathcal{P}_{\mu\nu}^{(2)\alpha\beta}}{-k^2 \left(1+ \kappa^2 \left(\frac{c_2(\mu)}{2} + 2 c_3(\mu) \right) k^2 - \kappa^2 \left(\beta/2 + 2 \gamma \right) k^2 \ln \left(-k^2/\mu^2\right) \right)} \\
&+ \frac{\mathcal{P}_{\mu\nu}^{(0)\alpha\beta}}{- k^2 \left(1 + \kappa^2 \left( -3 c_1(\mu) - c_2(\mu) - c_3(\mu) \right)  k^2 - \kappa^2 \left( -3 \alpha - \beta - \gamma \right)  k^2 \ln \left(-k^2/\mu^2\right) \right)} \ \ .
\end{align}
We decompose the trace-reduced metric perturbations (in harmonic gauge) as follows
\footnote{Note that the sum $\mathcal{P}^{(2)} + \mathcal{P}^{(0)} = \mathbbm{1}$ when it acts on symmetric tensors satisfying the harmonic gauge.}
\begin{align}
\bar{h}_{\mu\nu} = \bar{h}_{\mu\nu}^{(2)} + \bar{h}_{\mu\nu}^{(0)}, \quad \bar{h}_{\mu\nu}^{(2)} := \mathcal{P}_{\mu\nu}^{(2)\alpha\beta} \bar{h}_{\alpha\beta}, \quad \bar{h}_{\mu\nu}^{(0)} := \mathcal{P}_{\mu\nu}^{(0)\alpha\beta} \bar{h}_{\alpha\beta} \ \ .
\end{align}
We focus on the spin-2 sector and separate out the general relativity piece by re-writing the denominator in Eq.~(\ref{nonlocalprop})
\begin{align}
\nonumber
&\frac{1}{k^2 \left(1 + \kappa^2 \left(\frac{c_2(\mu)}{2} + 2 c_3(\mu) \right) k^2 - \kappa^2 \left(\beta/2 + 2 \gamma \right) k^2 \ln \left(-k^2/\mu^2\right) \right)} \\
&= \frac{1}{k^2} - \frac{ \kappa^2 \left(\frac{c_2(\mu)}{2} + 2 c_3(\mu) \right) - \kappa^2 (\beta/2 + 2\gamma) \ln \left(-k^2/ \mu^2\right)}{\left(1 + \kappa^2 \left(\frac{c_2(\mu)}{2} + 2 c_3(\mu) \right) k^2 - \kappa^2 \left(\beta/2 + 2 \gamma \right) k^2 \ln \left(-k^2/\mu^2\right) \right)} \ \ .
\end{align}
This way the spin-2 sector reads
\begin{align}
\bar{h}^{(2)}_{ij}(\omega,\vec{x}) = \bar{h}^{(2)\text{GR}}_{ij}(\omega,\vec{x}) + \bar{h}^{(2)\text{m}}_{ij}(\omega,\vec{x}),
\end{align}
where the massive spin-2 piece is now transparent. Working in the far-zone, we have
\begin{eqnarray}\label{nonlocalspin2}
\bar{h}^{(2)\text{m}}_{ij}(\omega,\vec{x}) &=& - (16 \pi G \kappa^2) I_{\ij}(\omega) \times \\ \nonumber && \int \frac{k^2 dk d\Omega_k}{(2\pi)^3} e^{i \vec{k} \cdot \vec{x}} \frac{\left(\frac{c_2(\mu)}{2} + 2 c_3(\mu) \right) -  (\beta/2 + 2\gamma) \ln \left(-k^2/ \mu^2\right)}{\left(1 + \kappa^2 \left(\frac{c_2(\mu)}{2} + 2 c_3(\mu) \right) k^2 - \kappa^2 \left(\beta/2 + 2 \gamma \right) k^2 \ln \left(-k^2/\mu^2\right) \right)},
\end{eqnarray}
where $I_{\ij}(\omega)$ is given in Eq.~(\ref{quadtensor}) and we work temporarily in a mixed frequency-position representation. Compared to Eq.~(\ref{resultsemifinal}), we observe that the non-analyticity has turned the denominator into a transcendental function which is infinitely-valued. A careful investigation of the latter is essential to understand the physical content of the result. The angular integrals in Eq.~(\ref{nonlocalspin2}) are readily performed
\begin{align}\label{kintegral}
\nonumber
&\bar{h}^{(2)\text{m}}_{ij}(\omega,\vec{x}) = (16 \pi G \kappa^2) I_{\ij}(\omega)  \left( \frac{1}{8 \pi^2 r} \right)  \times \\
& \frac{d}{dr} \int_{-\infty}^{\infty} dk  \frac{(e^{ikr} + e^{-ikr}) \left[\left(\frac{c_2(\mu)}{2} + 2 c_3(\mu) \right) -  (\beta/2 + 2\gamma) \ln \left((k^2 - \omega^2)/ \mu^2\right)\right]}{\left(1 + \kappa^2 \left(\frac{c_2(\mu)}{2} + 2 c_3(\mu) \right) (\omega^2 - k^2) - \kappa^2 \left(\beta/2 + 2 \gamma \right) (\omega^2 - k^2) \ln \left((k^2 - \omega^2)/\mu^2\right) \right)} \ \ ,
\end{align}
where it is understood that $\omega \to \omega + i\epsilon$ in the integrand to enforce retarded boundary conditions. Similar to the previous section, we evaluate the above integral in the complex plane. The situation here is rather complicated because the logarithm is infinitely-valued. This causes the integrand in Eq.~(\ref{kintegral}) to possess infinitely many poles that appear on the various Riemann sheets of the logarithm. The values of the poles are compactly encoded in the Lambert-W function \cite{Calmet:2014gya,Calmet:2018uub}
\begin{align}\label{spin2fullmass}
\omega^2 - k^2 = m_2^2 := \frac{1}{\kappa^2 (\beta/2 + 2 \gamma) W \left(- \frac{ 2 \exp\left(\frac{-c_2(\mu) - 4 c_3(\mu)}{\beta + 4 \gamma} \right)}{\kappa^2 \mu^2 (\beta + 4 \gamma)}  \right) } \ \ .
\end{align}

This reproduces the result obtained in \cite{Calmet:2018qwg}. We see from table (\ref{coeff1}) that the combination ($\beta/2 + 2 \gamma$) is positive-definite for all massless particles, and thus the argument of the Lambert-W function in Eq.~(\ref{spin2fullmass}) is negative-definite. 

\begin{figure}
\center
\begin{tikzpicture}
\draw [-] (-4,0) -- (4,0) node [below left]  {$\Re\{k\}$};
\draw [-] (0,-1) -- (0,4) node [below left] {$\Im\{k\}$};
 \begin{scope}[thick,font=\scriptsize]
    \draw [-,line width=1pt]  (4,0) arc (0:4:4);
    \draw [-] (3.99026,0.279026) -- (1.11808,0.279026);
    \draw [-,line width=1pt] (1.11808,0.279026) arc (90:-90:-0.120974);
    \draw [-] (1.11808,0.520974) -- (3.99026,0.520974);
    \draw  [-,line width=1pt](3.99026,0.520974) arc (7.43854:76.9864:4);
    \draw [black,fill=black] (1.11808,0.4) circle (0.05);
    \draw [black,fill=black] (2,2) circle (0.05);
    \draw (-2,2) node[cross, black] {};
    \draw [black,-,line width=1pt] (1.11808,0.4) -- (4,0.4);
    \draw [->,line width=1pt] (0.868081,3.90467) arc (77.4659:140:4) ;
    \draw [-,line width=1pt]  (-3.06418,2.57115) arc (140:180:4) ;
    \draw [->] (-4,0) -- (-2,0);
    \draw [-] (-2,0) -- (0,0);
    \draw [->] (0,0) -- (2,0);
    \draw [-] (0,0) -- (4,0);
    \end{scope}
 \end{tikzpicture}
    \caption{This figure shows our choice of integration contour in the complex $k$-plane, which is relevant for the integral involving the positive exponential factor in Eq.~(\ref{kintegral}). The horizontal line denotes the branch-cut in the upper-half-plane. The cross (dot) denotes the relevant pole if $\text{sgn}(\zeta)$ is positive (negative).}
    \label{contour}
    \end{figure}
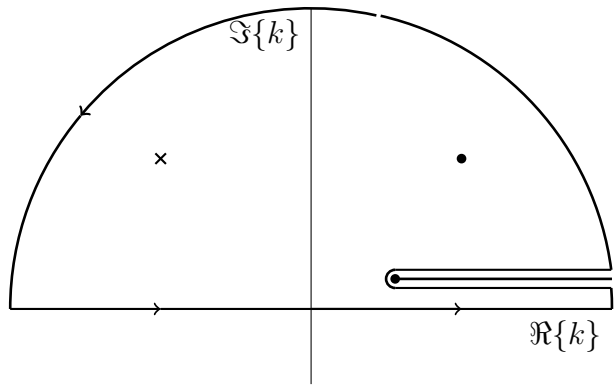

We will comment on the pole structure of Eq.~(\ref{kintegral}) as we proceed, but for now it suffices to pick a Riemann sheet in order to evaluate the integral. On each sheet, there is a single complex pole given any choice of the ultra-violet data, i.e. the Wilson coefficients and the renormalization scale \cite{Calmet:2017omb}. Let us treat in detail the integral involving the positive exponential in Eq.~(\ref{kintegral}), where our choice of the branch cut and integration contour is shown in Fig.~(\ref{contour}). Clearly, a generally complex solution to Eq.~(\ref{spin2fullmass}) introduces two poles which are mirror images of each other. Let us define two quantities
\begin{align}
\Omega := \omega^2 - \Re{m_2^2}, \quad \zeta := \Im{m_2^2} - \epsilon\, \text{sgn}(\omega) \ \ .
\end{align}
Notice that the sign of both $\Omega$ and $\zeta$ is not fixed at this stage. A direct computation yields 
\begin{align}\label{nonlocalpoles}
    k_\pm=\left\{
                \begin{array}{ll}
                 \pm \sqrt{\frac12 (\Omega^2 + \zeta^2)^{1/2} + \frac12 \Omega}\, \mp i \, \text{sgn}(\zeta) \sqrt{\frac12 (\Omega^2 + \zeta^2)^{1/2} - \frac12 \Omega} \ \ , \quad \Omega > 0 \\
                 \pm \sqrt{\frac12 (\Omega^2 + \zeta^2)^{1/2} - \frac12 | \Omega |}\, \mp i \, \text{sgn}(\zeta) \sqrt{\frac12 (\Omega^2 + \zeta^2)^{1/2} + \frac12 | \Omega |} \ \ , \quad \Omega < 0
                \end{array}
              \right.
\end{align}

Compared to Eq.~(\ref{localpoles}), we notice the important difference that the retarded $i\epsilon$-prescription does not play role in placing the poles because $\Im{m_2^2}$ is non-zero. For definiteness, let us focus on the case when $\Omega$ is positive. Since we close the contour in the upper-half-plane (cf. Fig.~(\ref{contour})), we only pick poles with positive imaginary part, and hence the contribution to the metric perturbations is Yukawa-suppressed. The same conclusion applies to the integral involving $e^{-ikr}$ as we close the contour in the lower-half-plane. The discontinuity across the branch cut cancel out in the final result and we are left with only the contribution from the residues.
\begin{align}\label{nonlocalinitial}
\nonumber
\bar{h}^{(2)\text{m}}_{ij}(t,\vec{x}) &= - 4 G \frac{ \mu  ( d \omega_s)^2 }{r}  \frac{ \frac{c_2(\mu)}{2} + 2 c_3(\mu) - (\beta/2 + 2\gamma) \ln \left(-m_2^2/ \mu^2 \right)}{ \frac{c_2(\mu)}{2} + 2 c_3(\mu) - (\beta/2 + 2\gamma) \ln \left(-m_2^2/ \mu^2 \right) + (\beta/2 + 2\gamma) } \times \\
& \exp \left( - r \sqrt{\frac12 (\Omega_s^2 + \zeta^2)^{1/2} - \frac12 \Omega_s} \right) \exp \left(- i r \,\text{sgn}(\zeta) \sqrt{\frac12 (\Omega_s^2 + \zeta^2)^{1/2} + \frac12 \Omega_s} \right)  \times \nonumber
 \\
& Q_{ij} (t,0;0) \ \ ,
\end{align}
where $\Omega_s := (2\omega_s)^2 - \Re{m_2^2}$. We immediately observe a problem with the above result, namely that the solution does not represent a propagating wave although $\Omega > 0$. Looking back at the local theory, we immediately realize that the reason for this is that the placement of the poles is not controlled by the sign of $\omega$ because $\Im{m_2^2}$ is non-zero. Moreover, the limit to the local theory ($\zeta \to 0$) does not exist given the structure of Eq.~(\ref{nonlocalinitial}).

In order to remedy this situation, we devise a new prescription for the poles in lieu of Eq.~(\ref{nonlocalpoles}). We first observe that the solutions to Eq.~(\ref{spin2fullmass}) come in conjugate pairs which appear on the mirror-symmetric Riemann sheets of the logarithm \cite{Calmet:2017omb}. Since one is free to pick a Riemann sheet on which to carry the contour integral, we demand the choice of the sheet to follow from the sign of the frequency. More precisely, let us say we picked a particular sheet and carried the integral for $\omega = 2\omega_s$, then the integral with $\omega = - 2\omega_s$ is to be evaluated on the mirror-symmetric sheet. We can summarize this prescription by staying on a single sheet but modifying equation Eq.~(\ref{nonlocalpoles}) to read
\begin{align}\label{causalnonlocalpoles}
    k_\pm=\left\{
                \begin{array}{ll}
                 \pm \sqrt{\frac12 (\Omega^2 + \zeta^2)^{1/2} + \frac12 \Omega}\, \mp i \, \text{sgn}(\omega) \sqrt{\frac12 (\Omega^2 + \zeta^2)^{1/2} - \frac12 \Omega} \ \ , \quad \Omega > 0 \\
                 \pm \sqrt{\frac12 (\Omega^2 + \zeta^2)^{1/2} - \frac12 | \Omega |}\, \mp i \, \text{sgn}(\omega) \sqrt{\frac12 (\Omega^2 + \zeta^2)^{1/2} + \frac12 | \Omega |} \ \ , \quad \Omega < 0
                \end{array}
              \right.
\end{align}

This prescription elegantly yields the desired behavior we are after. Let us also take the limit that the Wilson coefficients are large compared to $(\beta,\gamma)$\footnote{This limit gets rid of the prefactor appearing on the first line of Eq.~(\ref{nonlocalinitial}). Therefore, strictly speaking Eq.~(\ref{nonlocalfinal}) is correct up to corrections $\mathcal{O}\left((\beta + 4\gamma)/(c_2 + 4c_3)\right)$.}, hence we arrive at the radiation field 
\begin{align}\label{nonlocalfinal}
\bar{h}^{(2)\text{m}}_{ij}(t,\vec{x}) = - 4 G \frac{ \mu  ( d \omega_s)^2 }{r}  \exp \left( - r \sqrt{\frac12 (\Omega_s^2 + \zeta^2)^{1/2} - \frac12 \Omega_s} \right) Q_{ij} \left(t,r; m_{\text{eff}}^2\right) \ \ ,
\end{align}
where the effective mass of the wave is
\begin{align}\label{effmass}
m_{\text{eff}}^2 := (2\omega_s)^2 - \frac12 (\Omega_s^2 + \zeta^2)^{1/2} - \frac12 \Omega_s \ \ .
\end{align}

Eqs.~(\ref{nonlocalfinal}) and (\ref{effmass}) furnish the main results of our analysis in this section. Although we obtained Eq.~(\ref{nonlocalfinal}) for $\Omega_s > 0$, the corresponding result for $\Omega_s < 0 $ could readily be obtained using Eq.~(\ref{causalnonlocalpoles}). Thanks to our new prescription in Eq.~(\ref{causalnonlocalpoles}), the limit to the local theory ($\Im{m_2^2} \to 0$) exists and is manifest in our final result. As expected, Eq.~(\ref{nonlocalfinal}) represents a massive spherical wave albeit the amplitude is Yukawa suppressed due to the unavoidable imaginary part of the poles. Most importantly, the effective mass in Eq.~(\ref{effmass}) determines the speed of propagation of the wave. Finally, it is important to note that we did not place any restrictions regarding the signs and values of $\Re{m_2^2}$ and $\Im{m_2^2}$. From a phenomenological standpoint, it is crucial that the wave is sub-luminal, i.e. a positive-definite $m_{\text{eff}}^2$, which requires
\begin{align}
0 < \Re{m_2^2} \leq (2\omega_s)^2, \quad  \frac{\sqrt{\frac12 (\Omega_s^2 + \zeta^2)^{1/2} + \frac12 \Omega_s}}{2\omega_s} \leq 1 \ \ .
\end{align}

The calculation of the emitted power is complicated by the fact that the mass of the massive spin-2 field is now complex due to the non-local part of the action. A complex mass implies that this field has a width \cite{Calmet:2014gya} and a width cannot be implemented in a simple way in the Lagrangian. The calculation of the energy-momentum tensor $T_{\mu\nu}$ required to calculate the emitted power of a binary system into that mode is thus more complicated than in the local theory case. A standard way to introduce a width in a Lagrangian consists in including the interactions between the particle under consideration and its decaying product. It is clear that in the case, it will be an high order effect since we are working at second order in curvature and we can thus ignore the imaginary part of the mass. We thus recover the energy loss calculated in the previous section 
\begin{align}
\frac{dE_{\text{GW}}}{dt}  = \frac{32 G \mu^2 d^4 \omega_s^6}{5} \left( 1 + \theta(2\omega_s - \Re m_2) \right) + \mathcal{O}\left(\frac{\Re m_2}{\omega_s}\right) \ \ .
\end{align}  
where as before the first term is the power lost in the massless gravitational mode while the second term represents the power lost in the massive spin-2 mode \cite{Calmet:2018qwg}. This result was derived in \cite{Calmet:2018qwg}.

\section{Quantum non-locality: perturbative treatment}\label{pert}
While in the previous sections we studied effects at order $G$, i.e., the effects of the same strength as that of the standard general relativity gravitational wave solution, we now turn our attention to genuine quantum gravitational corrections to the general relativity wave solution which appear at order $G^2$. These corrections are the analogue of the long-distance corrections to the Newtonian potential, i.e. Eq.~(\ref{newtonian}), that have been derived in \cite{BjerrumBohr:2002ks,BjerrumBohr:2002kt}. To this aim, we look for a solution to Eq.~(\ref{NLEOM}) perturbatively close to general relativity
\begin{align}
\bar{h}_{\mu\nu} = \bar{h}^{\text{GR}}_{\mu\nu} + \mathfrak{h}_{\mu\nu}
\end{align}
where $\mathfrak{h}_{\mu\nu}$ comprises a {\em long-distance} correction to general relativity. Plugging this ansatz back in the equations of motion yields
\begin{align}\label{perteom}
\Box \mathfrak{h}_{ij} - \kappa^2 \left( \frac{c_2(\mu)}{2} - 2 c_3(\mu)  \right) \Box^2  \bar{h}^{\text{GR}}_{ij}+ \kappa^2 \left(\frac{\beta}{2} + 2\gamma \right) \Box^2 \frak{L}(\bar{h}^{\text{GR}}_{ij}) = 0 \ \ ,
\end{align}
where we have used the leading-order equation $\Box \bar{h}^{\text{GR}}_{\mu\nu} = -16 \pi G T_{\mu\nu}$. In our current approach the local pieces drop out, i.e. the middle term in Eq.~(\ref{perteom}), because away from the source we have that $\Box \bar{h}^{\text{GR}}_{\mu\nu} = 0$.  For the general relativity solution, we use the quadrupole formula
\begin{align}
\bar{h}^{\text{GR}}_{ij} = 4 G \frac{\mu  ( d \omega_s)^2}{r} Q_{ij}(t,r;0) \ \ .
\end{align}
We can simplify Eq.~(\ref{perteom}) if we commute one factor of the d'Alembertian past the logarithm in Eq.~(\ref{perteom}). The homogenous solution of $\mathfrak{h}_{\mu\nu}$ is set to zero and so we end up with
\begin{align}
\mathfrak{h}_{ij}  =  \frac{\kappa^4}{4} \left(\beta + 4\gamma \right) \mu  ( d \omega_s)^2 \, \frak{L}(\delta^{(3)}(\vec{x}) Q_{ij}(t,r;0)) \ \ .
\end{align}
At this point, the exact expression of $\frak{L}(x-x^\prime)$ derived in Eq.~(\ref{distribfinal}) is employed. The integral is quite involved, but we find it instructive to show some details that help illuminate the properties of the non-local distribution. Let us focus on a single component of the correction, say $\mathfrak{h}_{xx}$. The delta function allows us to integrate freely over spatial coordinates
\begin{eqnarray}
\mathfrak{h}_{xx} &=&  \frac{\kappa^4 \left(\beta + 4 \gamma \right)\mu  ( d \omega_s)^2}{4}  \times
\\ && \nonumber
 \lim_{\delta \to 0} \int dt^\prime \left[ \frac{i}{\pi^2} \left(\frac{ \Theta(t-t^\prime) \Theta((t -t^\prime)^2 - r^2)}{((t -t^\prime)^2 - r^2 + i\delta)^2} - \frac{ \Theta(t-t^\prime) \Theta((t -t^\prime)^2 - r^2)}{((t -t^\prime)^2 - r^2 - i\delta)^2}\right) \right] \cos(2\omega_s t^\prime).
\end{eqnarray}
Now the remaining integral is readily performed in the complex plane. Writing the cosine function in terms of complex exponentials, we close the contour appropriately. The step function $\Theta(t-t^\prime)$ picks up the causal pole and one ends up with manifestly real solutions
\begin{align}\label{pertsol}
\mathfrak{h}_{xx} = -\mathfrak{h}_{yy} &= \frac{\kappa^4 \left(\beta + 4 \gamma \right)\mu  ( d \omega_s)^2}{8\pi r^2} \left(2\omega_s  \sin(2 \omega_s t_r) - \frac{1}{r} \cos(2 \omega_s t_r) \right), \\
\mathfrak{h}_{xy} = \mathfrak{h}_{yx} &= -\frac{\kappa^4 \left(\beta + 4 \gamma \right)\mu  ( d \omega_s)^2}{8\pi r^2} \left(\frac{1}{r} \sin(2 \omega_s t_r) + 2\omega_s  \cos(2 \omega_s t_r) \right),
\end{align}
where $t_r := t- r$ is the retarded time. As we advertised, the above result represents a traveling massless wave, but with the far-field falling faster than the typical $1/r$ behavior of general relativity. A final comment is in place: the corrections in Eq.~(\ref{pertsol}) do not affect the radiated power since the field falls off faster than $1/r$. Since we are working perturbatively in $G$, the rate of energy loss is to be computed using the same expression in general relativity. Clearly as the power is obtained by averaging the energy flux over a sphere situated at infinity, any component in the wave solution that decays faster than $1/r$ does not contribute to the emitted power. This is not surprising, as here, the only degree of freedom involved that can carry energy is the massless spin-2 mode of general relativity. While the emitted power into massless gravitational waves is not corrected by quantum gravity at order $G^2$,  the strain which is given by
\begin{align}
h(t)=D^{\mu\nu}\bar{h}_{\mu\nu} = D^{\mu\nu}\bar{h}^{\text{GR}}_{\mu\nu} +D^{\mu\nu} \mathfrak{h}_{\mu\nu},
\end{align}
where $D^{\mu\nu}$ is the detector tensor, receives a quantum gravitational correction at this order.

\section{Conclusions}\label{Concs}

In this paper we worked within the effective theory approach to quantum gravity which enables model independent calculations at energies below the Planck mass.  The long-distance limit of quantum gravity is well described by the effective field theory framework. The advances in infrared quantum gravity opens the door to investigate a wide variety of gravitational observables. Using these now well established techniques, we reconsidered the question of quantum gravitational corrections to the emission of gravitational waves by a astrophysical binary system.

In this work we focused on the gravitational waveform emitted by a binary system during the inspiral phase. For completeness, we first revisited the production of massive spin-2 modes predicted by quantum gravity. We have then calculated the leading order quantum gravitational correction to the classical quadrupole radiation formula which appears at second order in Newton's constant. This is a genuine quantum gravitational prediction which is model independent. Clearly this is a  small effect which is unlikely to be relevant for any foreseeable gravitational wave experiment. However, this result is important as it demonstrates that quantum gravitational calculations are possible when using well established effective field theoretical techniques. This prediction of quantum gravity is model independent.  As expected, the emitted power into massless gravitational waves is not corrected by quantum gravity at order $G^2$. However, we have found that the strain receives a quantum gravitational correction at order $G^2$.


\bigskip{}

{\it Acknowledgments:}
The work of XC and BKE is supported in part by the Science and Technology Facilities Council (grant number ST/P000819/1). XC is very grateful to MITP for their generous hospitality during the academic year 2017/2018. SM is supported by a Chancellor's International Research Fellowship of the University of Sussex.

\section{Appendix}\label{app}

In this appendix, we derive the distribution $\mathfrak{L}(x-x^\prime)$ which formally reads
\begin{align}\label{kernelapp}
\frak{L}(x-x') = \int \frac{d^4 p}{(2\pi)^4}  e^{-i p \cdot (x-x')} \log{\left(\frac{-p^2}{\mu^2}\right)}  \ \ .
\end{align}
As it stands, the above integral is meaningless without specifying a boundary condition. To ensure causality, we impose retarded boundary conditions by writing $p^0 \to p^0 + i \epsilon$. In fact, this is not an {\em ad hoc} prescription. It was explicitly shown in \cite{Donoghue:2014yha} that using the in-in formalism to compute the effective action automatically yields a causal non-local distribution. Although ref. \cite{Donoghue:2014yha} was concerned with the time-dependent case, the conclusion is clear that in-in field theory guarantees the causal behavior of the equations of motion. 

We start by expressing the logarithm as follows
\begin{align}
\log{\left(\frac{-p^2}{\mu^2}\right)} = - \int_0^\infty dm^2 \left( \frac{1}{-p^2 + m^2} - \frac{1}{\mu^2 + m^2} \right) \ \ .
\end{align}
Notice that each integral diverges separately in such a way that the sum is finite. We have to introduce an explicit regulator, thus when we plug back in Eq.~(\ref{kernelapp})
\begin{align}\label{retnl}
\frak{L}(x-x') = \lim_{\delta \to 0} \left [\int_0^\infty \,dm^2 \int \frac{d^4 p}{(2\pi)^4} e^{-i p \cdot (x-x')} \frac{e^{-\delta \sqrt{\vec{p}^2 + m^2}}}{(p^0 + i \epsilon)^2-\vec{p}^2 - m^2} - \delta^{(4)}(x-x^\prime) \ln (\delta \mu)^2 \right] \ \ .
\end{align} 
As per usual, the integral over $p^0$ is readily performed and the poles are situated at
\begin{align}
p^0 = \pm \sqrt{\vec{p}^2 + m^2} - i\epsilon
\end{align}
which forces the integral to vanish if $x$ and $x^\prime$ are spacelike separated as one desires. Hence,
\begin{align}
\frak{L}(x-x') = \Theta(t-t^\prime) \Theta((x -x')^2) \lim_{\delta \to 0} \int_0^\infty \,dm^2 \int \frac{d^3p}{(2\pi)^3} e^{+i \vec{p} \cdot (\vec{x}-\vec{x}')} e^{-\delta \omega_p} \frac{\sin (\omega_p \Delta t)}{-\omega_p}
\end{align}
where $\omega_p := \sqrt{\vec{p}^2 + m^2}$ and $\Delta t := t-t^\prime$. Now the mass integral is easily done
\begin{align}
\frak{L}(x-x') = - \Theta(t-t^\prime) \Theta((x -x')^2) \lim_{\delta \to 0} \int \frac{d^3p}{(2\pi)^3} e^{+i \vec{p} \cdot (\vec{x}-\vec{x}')}\left(\frac{e^{i p (\Delta t + i\delta)}}{\Delta t + i \delta} + \frac{e^{-i p (\Delta t - i\delta)}}{\Delta t - i \delta} \right) \ \ .
\end{align}
The rest of the integral is elementary and yields a distribution, which is both Lorentz-invariant and retarded
\begin{eqnarray}\label{distribfinal}
\frak{L}(x-x') &=& \lim_{\delta \to 0} \Bigg[ \frac{i}{\pi^2} \left(\frac{ \Theta(t-t') \Theta((x -x')^2)}{((t -t^\prime +i \delta)^2 - (\vec{x}-\vec{x}')^2)^2} - \frac{ \Theta(t-t') \Theta((x -x')^2)}{((t -t^\prime - i \delta)^2 - (\vec{x}-\vec{x}')^2)^2}\right)  \nonumber \\   && - \delta^{(4)}(x-x^\prime) \ln (\delta \mu)^2 \Bigg] \ \ .
\end{eqnarray}
As we can see, this function has support only on the past light cone, which is as we expected. As a sanity check, this can also be seen to reduce to the cosmological expression found in \cite{Donoghue:2014yha,Fro2012} when we integrate out $d^3 x$.


\bigskip{}

 \end{document}